\documentclass[12pt]{article}
\let\section=\subsection     \let\subsection=\subsubsection                %%
\usepackage{graphicx}
\def\lesssim{\mathrel{\hbox{\rlap{\hbox{\lower4pt\hbox{$\sim$}}}\hbox{$<$}}}}
\begin{document}
\begin{center}
   {\large \bf $e^+e^-$ pair production from nucleon targets in the}\\[2mm]
   {\large \bf resonance region}\\[5mm]
    Madeleine Soyeur$\,^1$ and  Matthias F. M. Lutz$\,^2$\\[5mm]
   {\small \it $^1$DAPNIA/SPhN,
CEA/Saclay, F-91191 Gif-sur-Yvette CEDEX, France\\
$^2$GSI, Planckstrasse 1,  D-64291 Darmstadt, Germany\\
Institut f\"ur Kernphysik, TU Darmstadt,
   D-64289 Darmstadt, Germany
\\[8mm] }
\end{center}

\begin{abstract}\noindent
We present consistent theoretical descriptions of the $\pi\, N \rightarrow e^+e^- N$
and $\gamma\, N \rightarrow e^+e^- N$ reactions on proton and neutron targets for
total center of mass energies $\sqrt s$ ranging between $1.50$ GeV and $1.75$ GeV.
These reactions are complementary to study the coupling of low-lying
baryon resonances to vector meson-nucleon channels. We show in particular
how the resonant structure of the amplitudes for both processes gene\-rates
specific and large quantum interferences between $\rho$- and $\omega$-meson
decays into $e^+e^-$ pairs.
Data on the $\pi\, N \rightarrow e^+e^- N$
and $\gamma\, N \rightarrow e^+e^- N$ reactions
 are expected in the near future from
the HADES program at GSI and from dilepton studies with CLAS
at JLab.
\end{abstract}

\section{Introduction}
Low-lying resonances of masses $\lesssim\,$1.7 GeV do not decay
into a vector meson ($V=\rho,\omega$)
and a nucleon because there is no phase space for such decays
(except far on the resonance tails). The effective transition couplings of the
lowest-lying baryon excitations to vector field-nucleon final states are nevertheless
quantities of much significance to characterize the structure of baryons and
the propagation of vector mesons in the nuclear medium \cite{Riska1,Lutz1}.
These couplings can be best accessed through studies
of reactions in which baryon resonances are excited and decay
into a time-like photon (materializing subsequently into an
$e^+e^-$ pair) and a nucleon. We consider two such processes,
$\pi\, N \rightarrow e^+e^- N$
and $\gamma\, N \rightarrow e^+e^- N$. The model \cite{Lutz1}
underlying the description of these reactions makes use of
the information available on pion-nucleon and photonucleon
processes in the kinematic range where their dynamics is largely determined
by the excitation of s-channel baryon resonances.

We discuss the processes building up the amplitudes
for the $\pi\, N \rightarrow e^+e^- N$
and $\gamma\, N \rightarrow e^+e^- N$ reactions
and the way they are calculated in Section 2.
A few numerical results are displayed in Section 3 and
a brief conclusion is given in Section 4.
The work outlined in this talk relies on a published article
\cite{Lutz2} and on a forthcoming paper \cite{Lutz3}.

\section{The $\pi\, N \rightarrow e^+e^- N$
and $\gamma\, N \rightarrow e^+e^- N$ amplitudes}

The graphs entering the calculation of the $\pi\, N \rightarrow e^+e^- N$
and $\gamma\, N \rightarrow e^+e^- N$ amplitudes are displayed in Figs. 1 and 2.

\vglue 0.2truecm
\begin{center}
   \includegraphics[width=7cm]{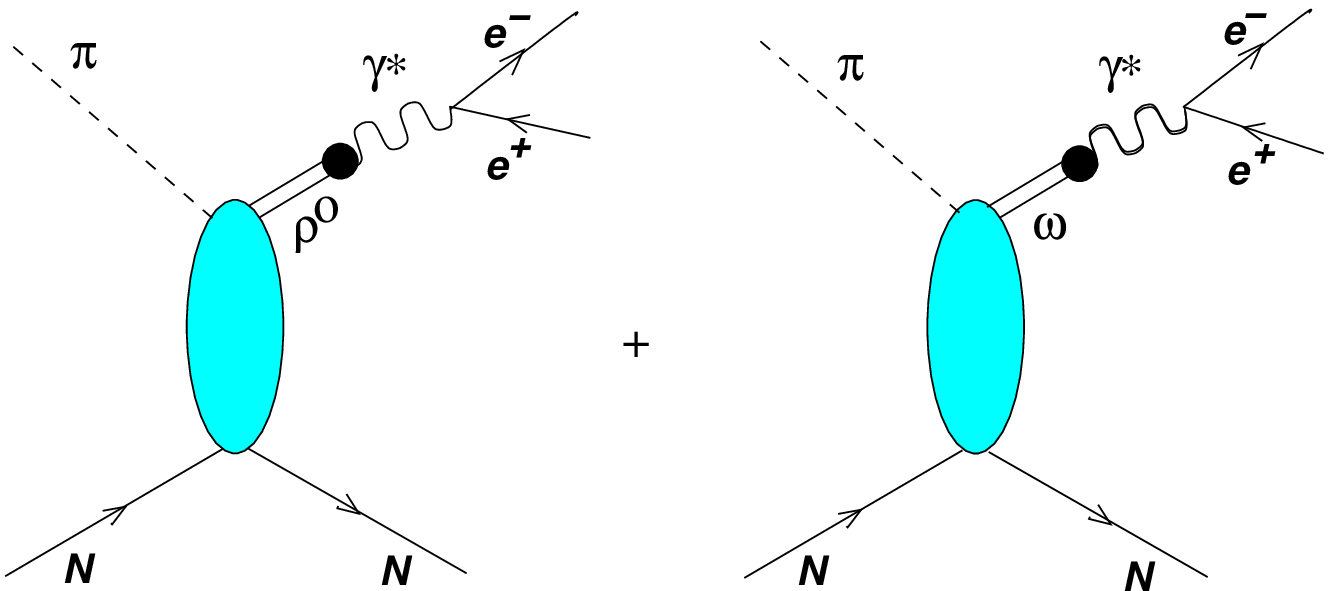}\\
   \vglue 0.1 truecm
   \parbox{14cm}
        {\centerline{\footnotesize
        Fig.~1: Diagrams contributing to the amplitude for the $\pi\, N \rightarrow e^+e^- N$ reaction.}}
\end{center}
\vskip 0.1truecm
\begin{center}
   \includegraphics[width=7cm]{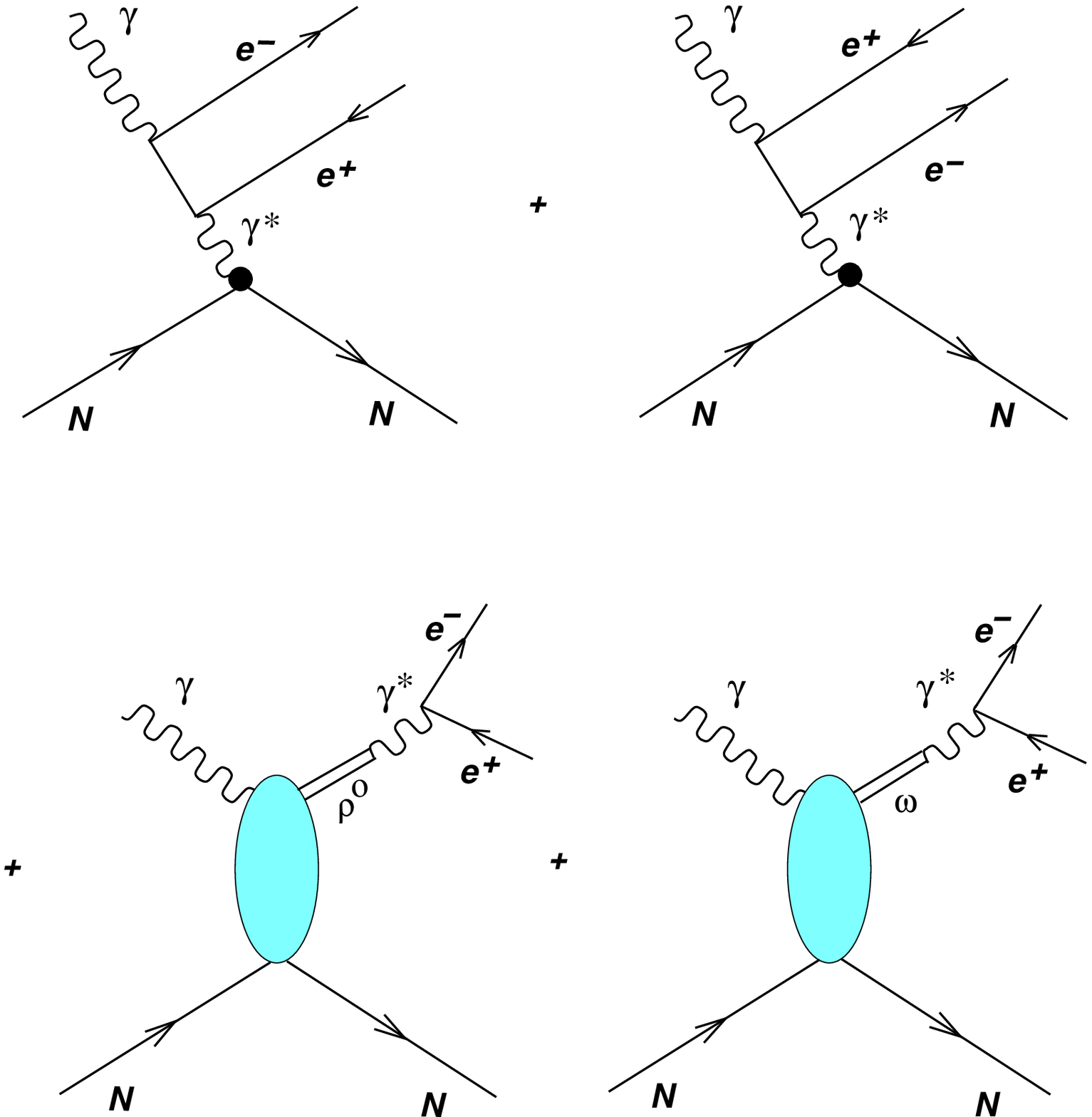}\\
   \vglue 0.1 truecm
   \parbox{14cm}
        {\centerline{\footnotesize
        Fig.~2: Diagrams contributing to the amplitude for the $\gamma\, N \rightarrow e^+e^- N$ reaction.}}
\end{center}
For the $\pi\, N \rightarrow e^+e^- N$ reaction, the amplitude consists of two terms
representing the production of a massive photon of isovector and isoscalar character
respectively. Using the Vector Meson Dominance assumption, these massive photons
can be related to the $\rho^0$- and $\omega$-meson fields \cite{Kroll}.
For the $\gamma\, N \rightarrow e^+e^- N$ reaction, there are four terms.
The first two diagrams are associated with the electromagnetic production of
$e^+e^-$ pairs (Bethe-Heitler processes). They depend on the (well-known) nucleon electromagnetic
form factors at low momentum transfers.
The last two diagrams represent the photoproduction of isovector
and isoscalar massive photons in the Vector Meson Dominance model.

The main dynamical quantities entering the calculation of the $\pi\, N \rightarrow e^+e^- N$
and $\gamma \, N \rightarrow e^+e^- N$ cross sections are therefore the (off-shell) $\pi\, N
\rightarrow \rho N$,
$\pi\, N \rightarrow \omega N$, $\gamma\, N \rightarrow \rho N$ and $\gamma \, N \rightarrow \omega N$
amplitudes. They are computed using the relativistic, chiral coupled-channel approach
of Ref. \cite{Lutz1}. This model offers a consistent picture of the $\pi\, N$ and
$\gamma\, N$ reactions and involves the $\pi N$, $\pi \Delta$, $\rho N$,
$\omega N$, $K \Lambda$, $K \Sigma$ and $\eta N$ hadronic channels. It is restricted to center
of mass energies ranging between $1.40$ GeV and $1.75$ GeV and des\-cribes vector meson-nucleon
channels below and very close to the vector meson threshold ($\sqrt s\simeq$1.72 GeV).
The vector meson and the nucleon in the final state are assumed to be
in relative S-wave. The Bethe-Salpeter kernel for the coupled-channel
system is constructed from an effective quasi-local meson-meson-baryon-baryon Lagrangian. The
fundamental fields are the photon, the mesons, the nucleon and the $\Delta(1232)$. The baryon
resonances which do not belong to the large N$_c$ groundstate multiplets are generated dynamically.
They are the N(1520), N(1535), N(1650), $\Delta$(1620) and $\Delta$(1700) resonances.
A generalized Vector Meson Dominance assumption is used to relate amplitudes involving photons to amplitudes
involving vector mesons. The effective Lagrangian parameters are fitted using all available data,
such as phase shifts, inelasticity parameters, pion photoproduction multipole amplitudes, inelastic
pion-nucleon cross sections. The quality of the fit is quite satisfactory for
all these quantities in the interval $1.40$ GeV $\leq \sqrt s \leq 1.75 \,$ GeV \cite{Lutz1}.

The presence of baryon resonances in this energy energy range is
reflected in the structure of the scattering amplitudes for vector
meson production. It is this particular structure that we are interested
in unravelling in $e^+e^-$ pair production processes. We illustrate
the resonant behavior of the vector
meson production amplitudes by showing in Fig. 3 the (projected)
amplitudes for the $\pi\, N \rightarrow \omega N$ process in the
S11 and D13 partial waves \cite{Lutz2}.
In the S${11}$ channel, the N(1535) and the N(1650) resonances
lead to peak structures in the ima\-ginary parts of the amplitudes.
The pion-induced $\omega$ production amplitudes in the D${13}$
channel reflect the strong coupling of the N(1520) re\-sonance to
the $\omega$N channel \cite{Lutz1}. The $\pi N
\rightarrow
\omega N$ amplitudes contain also significant contributions from
non-resonant, background terms.

The $\pi N \rightarrow e^+e^- N$ and $\gamma N \rightarrow e^+e^- N$
amplitudes involving intermediate vector mesons are calculated
assuming the specific Vector Meson Dominance form given by the
meson-photon interaction terms,
\begin{eqnarray}
{\mathcal L^{int}_{\gamma V}}\,&=&\, \frac {f_\rho} {2 M_\rho^2} F^{\mu \nu}\, \rho^0_{\mu \nu}
\,+\,\frac {f_\omega} {2 M_\omega^2} F^{\mu \nu}\, \omega_{\mu \nu},
\label{eq:e11}
\end{eqnarray}
\noindent
where the photon and vector meson field tensors are defined by
$F^{\mu \nu}\, =\,\partial^\mu A^\nu -\partial^\nu A^\mu$ and
$V^{\mu \nu}\, =\,\partial^\mu V^\nu -\partial^\nu V^\mu.$
The Bethe-Heitler terms are computed with phenomenological electromagnetic form factors
\cite{Lutz3}.
\begin{center}
   \includegraphics[width=9cm]{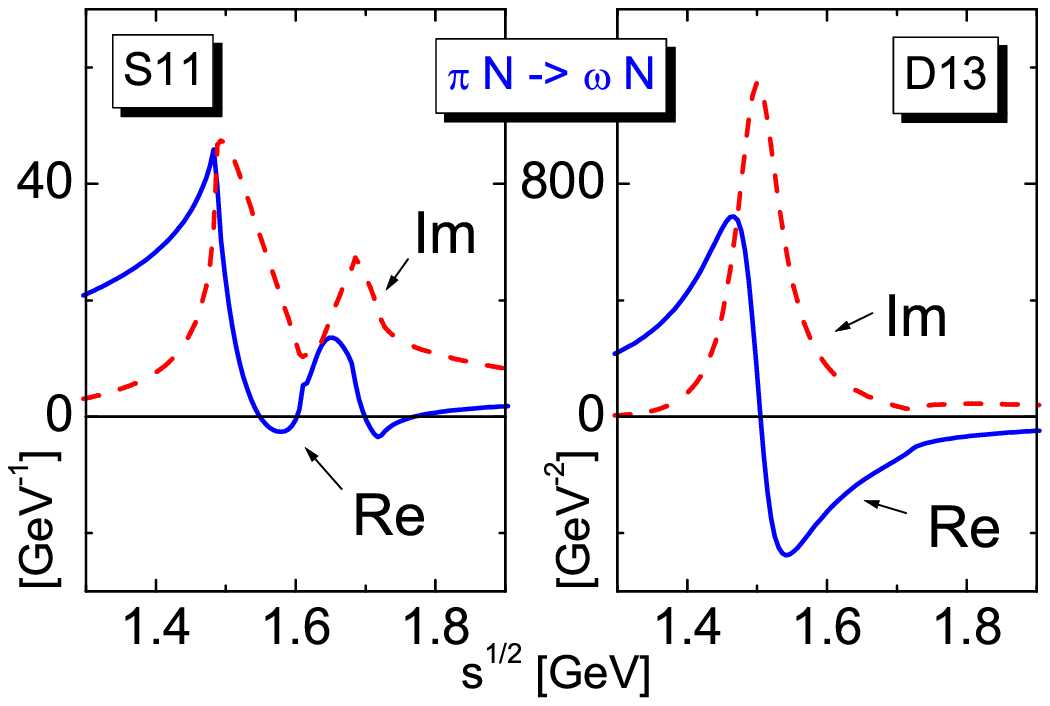}\\
   %\vglue 0.1 truecm
   \parbox{14cm}
        {\centerline{\footnotesize
        Fig.~3: Amplitudes for the $\pi\, N \rightarrow \omega N$ process in the
S11 and D13 partial waves \cite{Lutz2}.}}
\end{center}

\section{Numerical results}
A large set of numerical results for the $\pi\, N \rightarrow e^+e^- N$
and $\gamma\, N \rightarrow e^+e^- N$ reactions is presented in Refs.
\cite{Lutz2,Lutz3}. We display only a couple of illustrative figures in this paper.

Figs. 4 and 5 exhibit a very large quantum interference effect obtained
for the $\pi\, N \rightarrow e^+e^- N$ reaction in the lowest-lying resonance
region (N(1520), N(1535)) for $\sqrt s$=1.5 GeV.
The $\rho^0$-$\omega$ interference is
destructive for the $\pi^-p
\rightarrow e^+e^- n$ reaction and constructive for the $\pi^+n
\rightarrow e^+e^- p$ process.
The $\pi^-p
\rightarrow e^+e^- n$ differential cross section is extremely small
in the range of invariant masses considered in this calculation
(less than 10 nb GeV$^{-2}$). In contrast, the constructive
$\rho^0$-$\omega$ interference for the $\pi^+n \rightarrow e^+e^-
p$ reaction leads to a sizeable differential cross section (of the
order of 0.15 $\mu$b GeV$^{-2}$). This prediction is closely linked to the
resonant nature of the amplitudes and reflects the couplings of
both the N(1520) and N(1535) baryon
resonances to the
vector meson-nucleon channels. Data on these differential cross sections
 would provide strong constraints
on these couplings.
\begin{center}
   \includegraphics[width=7cm]{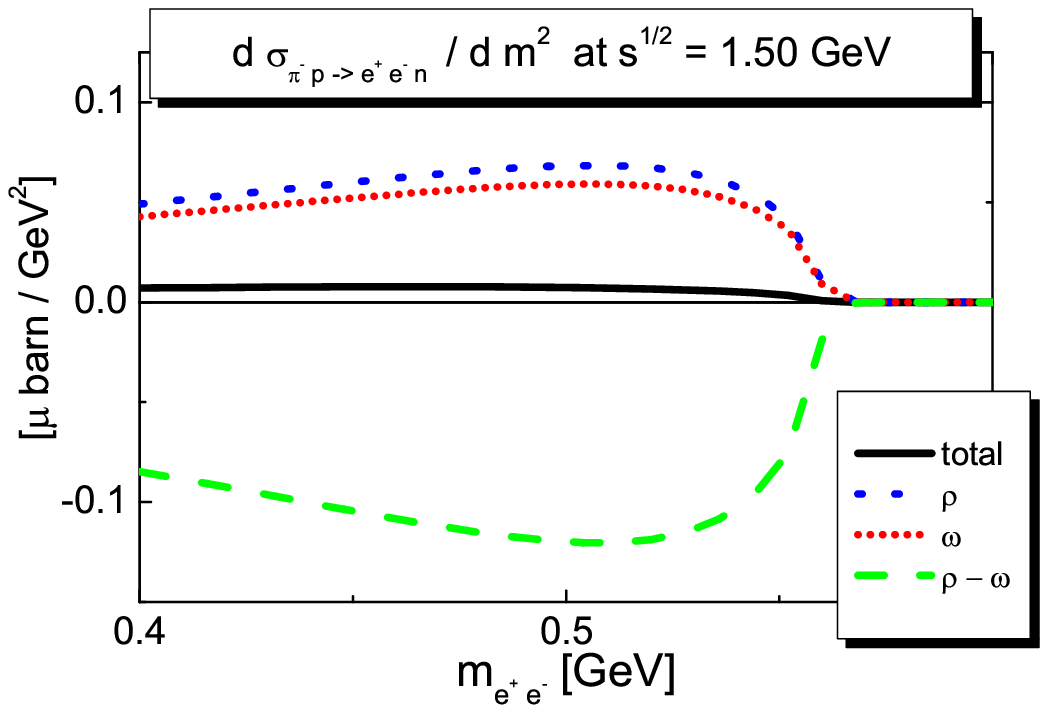}\\
   %\vglue 0.1 truecm
   \parbox{14cm}
        {{\footnotesize
        Fig.~4: Differential cross section for the
        $\pi^-p \rightarrow e^+e^- n$
reaction at $\sqrt s$=1.5 GeV
 \centerline{as function of the invariant mass of the $e^+e^-$
pair \cite{Lutz2}.}}}
\end{center}
\begin{center}
   \includegraphics[width=7cm]{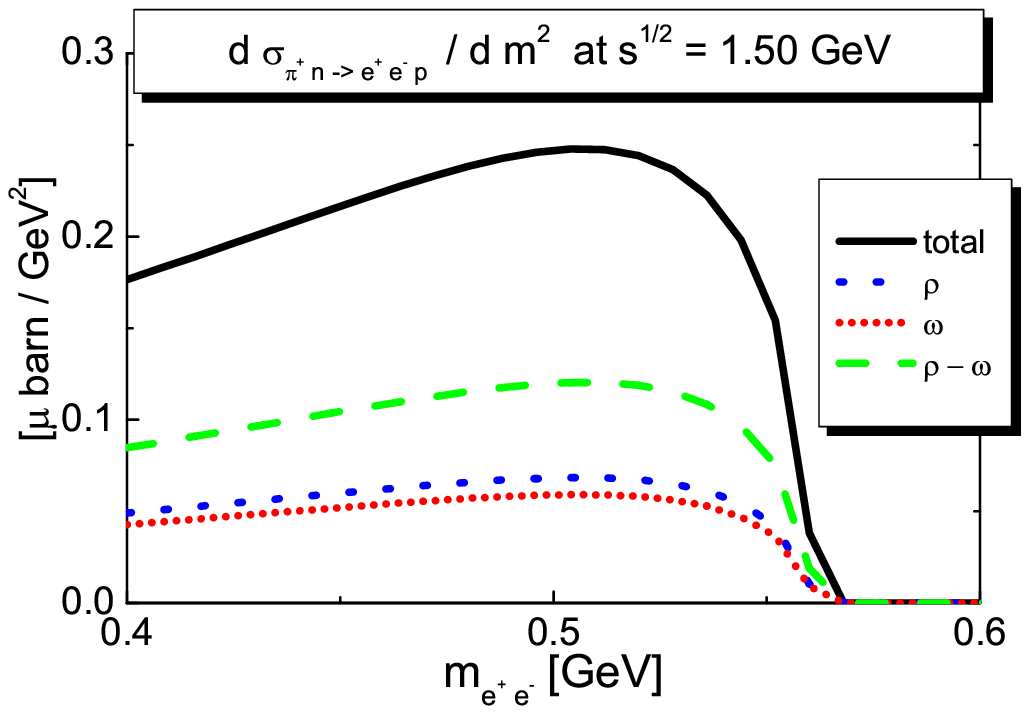}\\
   %\vglue 0.1 truecm
   \parbox{14cm}
        {\centerline{\footnotesize
        Fig.~5: Same as Fig. 4 for the $\pi^+n \rightarrow e^+e^- p$
reaction \cite{Lutz2}.}}
\end{center}
%\newpage
Fig. 6 illustrates the relative importance of $e^+e^-$ pair production through
Bethe-Heitler and vector meson production processes in the $\gamma p \rightarrow e^+e^- p$
reaction. The cross sections displayed are fully
integrated over both lepton three-momenta. There is no quantum interference between
Bethe-Heitler pairs and vector meson $e^+e^-$ decays in this case.
The right-hand side of Fig. 6 shows the $e^+e^-$ spectra produced in the
$\gamma p \rightarrow e^+e^- p$ reaction at $\sqrt s$=1.65  GeV.
For large $e^+e^-$ pair invariant masses, vector meson decays dominate over Bethe-Heitler
pair production. Computations of lepton pair angular distributions \cite{Lutz3} indicate
that Bethe-Heitler cross sections are strongly peaked at forward angles while vector meson
decays are much more isotropic. The latter can be best studied at backward angles.
The $\rho-\omega$ interference in the  $\gamma p \rightarrow e^+e^- p$ reaction is cons\-tructive
in both spin channels and dominated by $\rho$-meson production. This property
is related in particular to the coupling of the $\Delta$(1620) and $\Delta$(1700) resonances
to the $\rho$N channel. The $\gamma p \rightarrow e^+e^- p$ reaction appears as a much cleaner
process to study $\rho$-meson photoproduction than $\gamma p \rightarrow \pi^-\pi^+ p$, whose
cross section involves a large $\Delta \pi$ component in these kinematics. The left-hand side
of Fig. 6 shows that the $\gamma p \rightarrow e^+e^- p$ cross section increases very much
 above threshold and displays a very characteristic $\rho-\omega$ interference
pattern, while Bethe-Heitler processes represent only a small background.\par
\vglue 0.4 true cm \hskip 1truecm
   \includegraphics[width=7cm,angle=90]{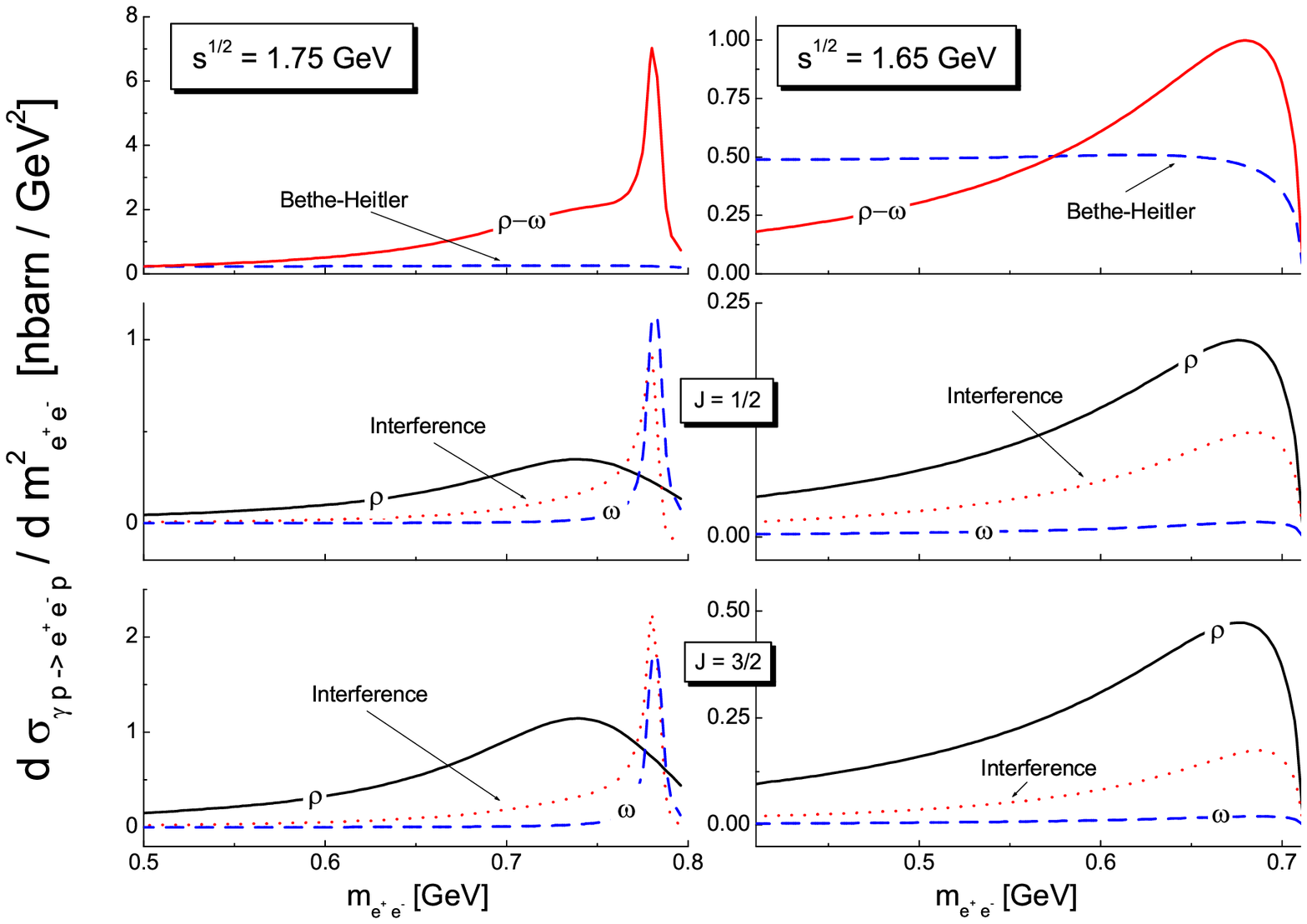}\\
\begin{center}
   \parbox{14cm}
        {\footnotesize
        Fig.~6: Differential cross section for the
        $\gamma p \rightarrow e^+e^- p$
reaction at $\sqrt s$=1.75  GeV and $\sqrt s$=1.65  GeV as function of the invariant mass of the $e^+e^-$
pair (preliminary result) \cite{Lutz3}.}
\end{center}
\section{Conclusion}
We have presented fragmentary results of an extensive theoretical study \cite{Lutz2,Lutz3}
of the $\pi\, N \rightarrow e^+e^- N$
and $\gamma\, N \rightarrow e^+e^- N$ reactions in the resonance region.
These reactions are sensitive to the coupling of low-lying
baryon resonances to vector meson-nucleon channels. These couplings imply
specific quantum interference patterns in the $e^+e^-$ decays of the virtual $\rho$- and $\omega$-mesons
produced in these reactions.
Data on these interferences would be extremely useful.


\begin{thebibliography}{99}
\itemsep=0cm
\bibitem{Riska1}
D.O. Riska and G.E. Brown, Nucl. Phys. A 679 (2001) 577.
\bibitem{Lutz1}
M.F.M. Lutz, Gy. Wolf and B. Friman, Nucl. Phys. A706 (2002) 431.
\bibitem{Lutz2}
M.F.M. Lutz, B. Friman, M. Soyeur, Nucl. Phys. A713 (2003) 97.
\bibitem{Lutz3}
M.F.M. Lutz, M. Soyeur, in preparation.
\bibitem{Kroll}
N.M. Kroll, T.D. Lee and B. Zumino, Phys. Rev. 157 (1967) 1376.
\end{thebibliography}
\end{document}